\documentstyle[11pt]{article}
\font\tenbf=cmbx10
\font\tenrm=cmr10
\font\tenit=cmti10
\font\elevenbf=cmbx10 scaled\magstep 1
\font\elevenrm=cmr10 scaled\magstep 1
\font\elevenit=cmti10 scaled\magstep 1

\renewenvironment{thebibliography}[1]
 { \elevenrm
   \begin{list}{\arabic{enumi}.}
    {\usecounter{enumi} \setlength{\parsep}{0pt}
     \setlength{\itemsep}{3pt} \settowidth{\labelwidth}{#1.}
     \sloppy
    }}{\end{list}}

\catcode`\@=11 
\def\lsim{\mathrel{\mathpalette\@versim<}}
\def\gsim{\mathrel{\mathpalette\@versim>}}
\def\@versim#1#2{\vcenter{\offinterlineskip
        \ialign{$\m@th#1\hfil##\hfil$\crcr#2\crcr\sim\crcr } }}
\catcode`\@=12 
\def\Tr{{\rm Tr\,}}
\def\D{{\partial}}

\begin{document}
\parindent=3pc
\baselineskip=10pt
\begin{center}{{\tenbf ELECTROWEAK SYMMETRY BREAKING\\
               \vglue 6pt
               AT THE SSC AND LHC}
\vglue 0.75cm
{\tenrm JONATHAN A. BAGGER \\}
\baselineskip=13pt
{\tenit Department of Physics and Astronomy\\}
\baselineskip=12pt
{\tenit The Johns Hopkins University\\}
{\tenit Baltimore, MD  21218\\}
}
\end{center}
\vglue 0.3cm
{\rightskip=3pc
 \leftskip=3pc
 \tenrm\baselineskip=12pt
 \noindent
In this talk we survey the SSC and LHC signals and backgrounds
for the physics of electroweak symmetry breaking.  We study the
process $pp \rightarrow WWX$ and compute the rate for the
``gold-plated'' signals $W^\pm \rightarrow \ell^\pm \nu$ and
$Z \rightarrow \ell^+\ell^-$ $(\ell = e,\mu)$ for a wide variety
of models.   We use a forward jet tag and central jet veto to
suppress the standard-model backgrounds.  In this way we estimate
the SSC and LHC sensitivities to the physics of electroweak symmetry
breaking.
\vglue 0.6cm}

{\elevenbf\noindent 1. Introduction}
\vglue 0.2cm
\baselineskip=14pt
\elevenrm

In the standard model of particle physics, electroweak
symmetry is broken by a the vacuum expectation value
of a Higgs particle $H$, whose mass is expected
to be less than a TeV.  To date, however, there is still
no experimental evidence -- either direct or indirect --
in favor of the Higgs.

One fact is certain:  New physics is needed to break the
electroweak symmetry.  If one uses the known particles
and computes the scattering amplitude for
longitudinally-polarized $W$'s, one finds that rate
the diverges with energy, and that (perturbative)
unitarity is violated below 2 TeV.  New physics
is necessary to unitarize the scattering amplitude.$^1$
In the standard model, the new physics is just the Higgs
particle $H$.

In this talk we will look beyond the Higgs
to study the SSC and LHC signals and backgrounds for
a wide variety of models that break electroweak symmetry and
unitarize the $W_LW_L$ scattering amplitude.  Each of
the models is completely consistent with all the data to
date (including that from the $Z$).  Together, they
indicate the range of new physics that might
be seen at the SSC or LHC.

\vglue 0.6cm
{\elevenbf\noindent 2. The Models}
\vglue 0.4cm

At present, all we know about electroweak symmetry breaking
is that $M_W \simeq M_Z \cos\theta$, which suggests that
the underlying physics respects a global
symmetry $G \supseteq SU(2)_L\times SU(2)_R$, spontaneously
broken to $H \supseteq SU(2)_V$.  In this talk we will examine
a wide variety of models consistent with this ``isospin''
symmetry.  The first major distinction is whether or not
a given model is resonant in the $W_LW_L$ channel.  If it is
resonant, the model can be classified by the spin and isospin
of the resonance.  If it is not, the analysis is more
subtle, and we shall see that all possibilities can be
described in term of two parameters.

\vglue 0.2cm
{\elevenit \noindent 2.1. Spin-zero, Isospin-zero Resonances}
\vglue 0.1cm

1) {\elevenit Standard Model.}
The standard model is the prototype of a theory with a
spin-zero, isospin-zero resonance.  The $W_LW_L$ scattering
amplitudes are unitarized by exchange of the Higgs particle
$H$.  The Higgs is contained in a complex scalar doublet,
$\Phi\ =\ (v + H) \exp(2i w^a \tau^a/v)$, whose four components
split into a triplet $w^a$ and a singlet $H$ under isospin.

The standard-model Higgs potential is invariant under an $SU(2)_L
\times SU(2)_R$ symmetry.  The vacuum expectation value $\langle\Phi\rangle
= v$ breaks the symmetry to the diagonal $SU(2)$.  In the perturbative
limit, it also gives mass to the Higgs.  For the purposes of this talk,
we will take $M_H = 1$ TeV.

2)  $O(2N).$
This model attempts to
describe the standard-model Higgs in the nonperturbative
domain.  In the perturbatively-coupled standard model,
the mass of the Higgs is proportional to the square root
of the scalar self-coupling $\lambda$.  Heavy Higgs
particles correspond to large values of $\lambda$.
For $M_H \gsim$ 1 TeV, naive perturbation theory breaks
down.

One way to explore the nonperturbative regime
is to exploit the isomorphism between $SU(2)_L \times SU(2)_R$
and $O(4)$.  Using a large-$N$ approximation, one can
solve the $O(2N)$ model for all values of $\lambda$,
to leading order in $1/N$.  The resulting scattering
amplitudes$^2$ can be parametrized by the scale $\Lambda$ of
the Landau pole.  Large values of $\Lambda$ correspond
to small couplings $\lambda$ and relatively light Higgs
particles. In contrast, small values of $\Lambda$ correspond
to large $\lambda$ and describe the nonperturbative regime.
In this talk we will take $\Lambda = 3$ TeV.

\vglue 0.2cm
{\elevenit \noindent 2.2. Spin-one, Isospin-one Resonances}
\vglue 0.1cm

1)  {\elevenit Vector.}
This model provides a relatively model-independent
description of the techni-rho resonance that arises in
most technicolor theories.  One can use the time-honored
techniques of chiral Lagrangians to construct the
coupling between the techni-rho and the Goldstone
bosons.$^{3,4}$  The basic fields are $\xi = \exp(i w^a
\tau^a/v)$ and a vector $\rho_\mu$, which transform
nonlinearly under $SU(2)_L \times SU(2)_R$.

\begin{table}[t]
\begin{center}\vspace{-.19in}
\begin{tabular}{ l l | l l}
\multicolumn{4}{l}{\tenrm Table~1. SSC cuts, tags and vetoes, by mode.}\\[2mm]
\hline\hline
$W^+ W^-$ Basic cuts & Tag and Veto & $Z Z$ Basic cuts & Tag only\\
\hline
 $| y_{\ell} | < 2.0 $  &
 $E_{tag} > 3.0\ {\rm TeV}$  &
 $| y_{\ell} | < 2.5 $  &
 $E_{tag} > 1.0\ {\rm TeV}$   \\
 $P_{T,\ell} > 100\ {\rm GeV}$  &
 $3.0 < \eta_{tag} < 5.0$  &
 $P_{T,\ell} > 40\ {\rm GeV}$  &
 $3.0 < \eta_{tag} < 5.0$   \\
 $\Delta P_{T,\ell\ell} > 200\ {\rm GeV}$  &
 $P_{T,tag} > 40\ {\rm GeV}$  &
 $P_{T,Z} > {1\over4} \sqrt{M^2_{ZZ} - 4 M^2_Z}$  &
 $P_{T,tag} > 40\ {\rm GeV}$   \\
 $\cos\phi_{\ell\ell} < -0.8$  &
 $P_{T,veto} > 60\ {\rm GeV}$  &
 $M_{ZZ} > 500\ {\rm GeV}$  & \\
 $M_{\ell\ell} > 250\ {\rm GeV}$  &
 $ | \eta_{veto} | < 3.0$ && \\[2mm]
\hline
$W^+ Z$ Basic cuts & Tag and Veto &
$W^+ W^+$ Basic cuts & Veto only \\
\hline
 $| y_{\ell} | < 2.5 $  &
 $E_{tag} > 2.0\ {\rm TeV}$  &
 $| y_{\ell} | < 2.0 $   &
 $P_{T,veto} > 60\ {\rm GeV}$  \\
 $P_{T,\ell} > 40\ {\rm GeV}$  &
 $3.0 < \eta_{tag} < 5.0$  &
 $P_{T,\ell} > 100\ {\rm GeV}$   &
 $ | \eta_{veto} | < 3.0$  \\
 $P_{T, miss} >  75\ {\rm GeV}$  &
 $P_{T,tag} > 40\ {\rm GeV}$  &
 $\Delta P_{T,\ell\ell} > 200\ {\rm GeV}$   & \\
 $P_{T,Z} > {1\over4} M_T{}^*$ &
 $P_{T,veto} > 60\ {\rm GeV}$  &
 $\cos\phi_{\ell\ell} < -0.8$  & \\
 $M_T > 500\ {\rm GeV}$ &
 $ | \eta_{veto} | < 3.0$  &
 $M_{\ell\ell} > 250\ {\rm GeV}$  & \\
\hline\hline
\multicolumn{2}{l}{{}$^*$  \tenrm $M_T$ is the cluster transverse
mass.}\\
\end{tabular}
\end{center}
\end{table}

For the processes of interest, the effective Lagrangian
depends on just two couplings, which we can take to
be the mass and the
width of the resonance.  In what follows we will choose
$M_\rho = 2.0$ TeV, $\Gamma_\rho = 700$ GeV and
$M_\rho = 2.5$ TeV, $\Gamma_\rho = 1300$ GeV.
These values preserves unitarity up to 3 TeV.

\vglue 0.2cm
{\elevenit \noindent 2.3. Nonresonant models}
\vglue 0.1cm

The final models we consider are
nonresonant at SSC energies.  In this case
the new physics contributes to the effective Lagrangian
in the form of higher-dimensional operators built
from the Goldstone fields.  To order $p^4$ in the
energy expansion, there are only three
operators that contribute to $W_LW_L$
scattering.$^5$  They are
\begin{equation}
{\cal L} \ = \ {v^2\over4}\,\Tr \D_\mu \Sigma
\D_\mu \Sigma^\dagger\ +\
{L_1\over 16\pi^2}\,\bigg(\Tr \D_\mu \Sigma
\D_\mu \Sigma^\dagger\bigg)^2\
 +\ {L_2\over 16\pi^2}\,\bigg(\Tr \D_\mu \Sigma
\D_\nu \Sigma^\dagger\bigg)^2\ .
\end{equation}
\noindent
The coefficients $L_1$ and $L_2$ contain all information
about the new physics.

The difficulty with this approach is that at SSC energies,
the scattering amplitudes violate unitarity between 1 and 2
TeV.  This is an indication that new physics is near, but
not necessarily within the
reach of the SSC. We choose to treat the uncertainties of
unitarization in two ways:

1)  {\elevenit LET CG.}
We take $L_1 = L_2 = 0$, and cut off the partial wave
amplitudes when they saturate the unitarity bound.$^1$

2)  {\elevenit Delay K.}
We take $L_1 = -0.26$ and $L_2 = 0.23$, a choice that
preserves unitarity up to 2 TeV.  Beyond that scale, we
unitarize the scattering amplitudes with a K-matrix.

\vglue 0.6cm
{\elevenbf\noindent 3. Signal and Backgrounds}
\vglue 0.4cm

In the rest of this talk we will focus on SSC and LHC signals
and backgrounds for the process $pp \rightarrow WWX$.  We will
concentrate on the ``gold-plated'' decays
$W^\pm \rightarrow \ell^\pm \nu$ and $Z \rightarrow \ell^+
\ell^-$, for $\ell = e,\mu$, in each of the final states
$W^+W^-$, $W^+ Z$, $ZZ$ and $W^+W^+$.

We will take the signal to be the process $pp \rightarrow
W_LW_LX$ because the longitudinal $W$'s couple most
strongly to the new physics.  We will take $pp \rightarrow
W_LW_TX$ and $pp \rightarrow W_TW_TX$ to be the background.
These processes are dominated by diagrams that do not depend
on the new physics, so we will
represent the background by the standard model with a light
Higgs (of mass 100 GeV).
The difference between this and the true background is
negligible at the energies we consider.

We will simplify our calculations by using the
equivalence theorem, which lets us replace the longitudinal
vector bosons by their corresponding would-be Goldstone
bosons.  We will also use the effective $W$ approximation
to connect the $W_LW_L$ subprocesses to the $pp$
initial state.

In the $W^+W^-$, $W^+ Z$ and $ZZ$ channels, the final
states of interest are dominated by glue-glue and
$q \bar q$ scattering.  We suppress these contributions
by requiring a tag on the forward jet$^6$ associated with
an initial-state $W$.
In the $W^+W^-$, $W^+ Z$ and $W^+W^+$ channels, there
is a residual background from top decay that we suppress
by requiring a central jet veto.$^7$  The combination
of a forward jet tag and central jet veto is very effective
in reducing the background in all charge channels.

The precise SSC cuts we use are summarized in Table 1.  (The
LHC cuts are the same except for the jet tags, which are
$E_{jet} > 2.0$, 1.5,  and 0.8 TeV in the $W^+W^-$,
$W^+Z$, and $ZZ$ channels, respectively.)
In all channels, the dominant residual background is transverse
electroweak, followed by $q \bar q$ annihilation and top decay.

Because we use the effective $W$ approximation for our signal,
we can only estimate the effects of the tag and veto.  Therefore
we have used the exact standard-model calculation with a 1 TeV
Higgs to derive efficiencies for the tag and veto.  These
efficiencies are then applied to the effective $W$ calculations
to estimate the rate for each signal.  The results for the
signals and backgrounds are collected in Table 2.

\begin{table}[t]
\begin{center}\vspace{-.19in}
\begin{tabular}{ l | c  c  c  c  c  c  c }
\multicolumn{8}{l}{\tenrm Table~2. Event rates per SSC/LHC-year, assuming $m_t
= 140$ GeV, $\sqrt{s} = 40/16$ TeV,}\\
\multicolumn{8}{l}{\tenrm \hspace{38pt}  and an annual luminosity of
$10^4/10^5$
pb$^{-1}$.}\\[2mm]
\hline\hline
$W^+ W^-$ & Bkgd. & SM 1.0 &  $O(2N)$ &  Vec 2.0 & Vec 2.5 & LET CG & Delay K
\\
\hline
$M_{\ell\ell} > 0.25$ & 9.1/13  & 59/74   & 26/30 &    12/9.7  & 10/8.9  &
12/11   & 9.7/8.9  \\
$M_{\ell\ell} > 0.5$  & 5.0/5.6   & 31/32   & 16/16 &    9.3/6.4 & 7.4/5.3 &
9.3/7.0 & 6.9/5.0  \\
$M_{\ell\ell} > 1.0$  & 0.9/0.7   & 2.0/1.2 & 1.5/0.8 &  3.6/1.8 & 2.6/1.2 &
2.9/1.3 & 2.5/0.9
\\[2mm]
\hline
$W^+ Z$ & Bkgd. & SM 1.0 &  $O(2N)$ &  Vec 2.0 & Vec 2.5 & LET CG & Delay K  \\
\hline
$M_T > 0.5$ & 2.5/2.3 & 1.3/1.0 &  1.5/1.1 &  9.6/4.8  &  6.2/3.2
& 5.4/3.1 & 5.5/2.9  \\
$M_T > 1.0$ & 0.9/0.4 & 0.6/0.3 &  0.8/0.4 &  8.2/3.4  &  4.8/1.9
& 4.0/1.7 & 4.3/1.7  \\
$M_T > 1.5$ & 0.3/0.1 & 0.2/0.1 &  0.3/0.1 &  5.9/2.1  &  3.4/1.1
& 2.5/0.8 & 2.9/0.9
\\[2mm]
\hline
$Z Z$ & Bkgd. & SM 1.0 &  $O(2N)$ &  Vec 2.0 & Vec 2.5 & LET CG & Delay K  \\
\hline
$M_{ZZ} > 0.5$ & 1.0/1.0 & 11/14  &  5.2/6.4&  1.1/1.4 & 1.5/1.7 &
2.5/2.5 &  1.5/1.8
\\
$M_{ZZ} > 1.0$ & 0.3/0.2 & 4.8/4.8 &  2.3/2.2 &  0.5/0.4 & 0.8/0.6 &
1.7/1.2 &  0.8/0.6
\\
$M_{ZZ} > 1.5$ & 0.1/0.0 & 0.6/0.4 &  0.5/0.3 &  0.1/0.1 & 0.3/0.2 &
0.9/0.5 &  0.3/0.2
\\[2mm]
\hline
$W^+ W^+$ & Bkgd. & SM 1.0 &  $O(2N)$ &  Vec 2.0 & Vec 2.5 & LET CG &  Delay K
\\
\hline
$M_{\ell\ell} > 0.25$ & 3.5/6.0 & 6.4/9.6 &  7.1/10 &  7.8/12 & 11/16 &
25/27 & 15/16  \\
$M_{\ell\ell} > 0.5$ & 1.9/2.1 & 3.8/4.6 &  4.5/5.2 &  4.5/6.0 & 7.2/8.8 &
20/18 & 11/9.6  \\
$M_{\ell\ell} > 1.0$ & 0.3/0.3 & 0.7/0.5 &  1.1/0.7 &  0.6/0.5 & 1.5/1.2 &
8.3/4.9 & 5.3/2.6  \\[2mm]
\hline\hline
\end{tabular}
\end{center}
\end{table}

\vglue 0.6cm
{\elevenbf\noindent 4. Discussion}
\vglue 0.4cm

The results in Table 2 summarize the outcome of our
study.  As expected, the signal rates are largest in
the resonant channels.  Note, however, that the
rates are all rather low.  The events are clean, but
the low rates will make it difficult to isolate
high-mass resonances.  We must be prepared for
a high-luminosity program at the SSC and LHC.

A second conclusion from Table 2 is that all channels are
necessary.  For example, isospin-zero resonances
give the best signal in the $W^+W^-$ and $ZZ$ channels, while
isospin-one resonances dominate the $W^+ Z$ channel.  The
nonresonant models tend to show up in the $W^+W^+$ final state,
so there is a complementarity between the different
channels.$^1$

A third conclusion is that we cannot cut corners.
Accurate background studies are crucial if we hope to
separate signal from background by simply counting rates.
We must also try to measure all decay modes
the $W$ and $Z$, including $Z \rightarrow \nu\bar\nu$ and
$W,Z \rightarrow jets$.  Finally, we must work to optimize the
cuts that are applied to each final state, with an eye to
increasing the signal/background ratio without affecting the
total rate.  All these considerations indicate that if
electroweak symmetry is dynamically broken,
SSC and LHC studies of electroweak symmetry breaking might
need a mature and long-term program before they give rise to
fruitful results.

\vglue 0.6cm
{\elevenbf\noindent 5. Acknowledgements}
\vglue 0.4cm

This work was supported in part by NSF grant PHY-90-96198.
I would like to thank my collaborators V.~Barger, K.~Cheung,
J.~Gunion, T.~Han, G.~Ladinsky, R.~Rosenfeld and C.-P.~Yuan.
I would also like to thank S.~Dawson and G.~Valencia for many
conversations on the effective Lagrangian approach
to electroweak symmetry breaking.

\vglue 0.6cm
{\elevenbf\noindent 6. References}
\vglue 0.4cm

\end{document}